\documentclass[12pt]{article}

\usepackage{times}
\usepackage{float}
\usepackage{soul}

\newenvironment{sciabstract}{%
\begin{quote} \bf}
{\end{quote}}

\title{Reconnection-Driven Energy Cascade in Magnetohydrodynamic Turbulence}

\author{Chuanfei Dong$^{1,2}$\thanks{Corresponding author. Email: dcfy@princeton.edu.}$^{~,\dag}$, Liang Wang$^{2,1}$\thanks{These authors contributed equally to this work.}, Yi-Min Huang$^{2,1}$,\\
Luca Comisso$^{3}$, Timothy A. Sandstrom$^{4}$, Amitava Bhattacharjee$^{1,2}$\\
\normalsize{$^{1}$Princeton Plasma Physics Laboratory, Princeton University, Princeton, NJ 08540, USA}\\
\normalsize{$^{2}$Department of Astrophysical Sciences, Princeton University, Princeton, NJ 08544, USA} \\
 \normalsize{$^{3}$Department of Astronomy and Columbia Astrophysics Laboratory,}\\
 \normalsize{Columbia University, New York, NY 10027, USA}\\
 \normalsize{$^{4}$NASA Ames Research Center, Mountain View, CA 94043, USA}
\\
}

\date{}

\usepackage{graphicx}
\usepackage{bm}
\usepackage{color}
\usepackage{amsmath}
\usepackage{amsfonts}
\usepackage{amssymb}
\usepackage{latexsym}
\usepackage{mathrsfs}
\usepackage[hidelinks,draft=false]{hyperref}
\usepackage{color}
\usepackage{wasysym}
\usepackage{textcomp}

\usepackage{changes} %

\begin{document}


\baselineskip24pt


\maketitle

\begin{sciabstract}
Magnetohydrodynamic turbulence regulates the transfer of energy from large to small scales in many astrophysical systems, including the solar atmosphere. We perform three-dimensional magnetohydrodynamic simulations with unprecedentedly large magnetic Reynolds number to reveal how rapid reconnection of magnetic field lines changes the classical paradigm of the turbulent energy cascade. By breaking elongated current sheets into chains of small magnetic flux ropes (or plasmoids), magnetic reconnection leads to a new range of turbulent energy cascade, where the rate of energy transfer is controlled by the growth rate of the plasmoids. As a consequence, the turbulent energy spectra steepen and attain a spectral index of -2.2 that is accompanied by changes in the anisotropy of turbulence eddies. The omnipresence of plasmoids and their consequences on, e.g., solar coronal heating, can be further explored with current and future spacecraft and telescopes.
\end{sciabstract}

\section*{INTRODUCTION}
Understanding the transfer of energy in magnetohydrodynamic (MHD) turbulence is crucial for tackling many outstanding astrophysical problems, such as solar and stellar coronal heating, star formation, cosmic-ray transport, and the interstellar medium evolution. For more than half a century, it has been widely accepted that the energy cascade in turbulent plasmas such as the Sun's atmosphere is controlled by magnetohydrodynamic wave interactions \cite{Iroshnikov1963,Kraichnan1965,GS95}. However, one essential feature of magnetohydrodynamic turbulence is the ubiquitous presence of sheets of intense electric current (known as current sheets), which are preferential locations for rapid breaking and reconnection of magnetic field lines \cite{ML86,Biskamp1989}, a fundamental physical process in magnetized plasmas whereby stored magnetic energy is converted into heat and kinetic energy of charged particles. It is yet an open question whether magnetic reconnection can significantly change the transfer of energy from large to small scales in a wide variety of astrophysical systems.

One of the most common approaches to investigating the effect of a physical process on the turbulent energy cascade is to study the associated energy spectra. It has been widely observed by different spacecraft and telescopes that turbulent energy spectra can break at either visco-resistive or kinetic scales \cite{Kiyani2015RSPTA,Abramenko2020,Kieokaew2021}. Recent analytic studies suggest that magnetic reconnection may also break the turbulent energy spectra and thus create a new range of energy transfer when the growth time scale of the magnetic flux ropes (or plasmoids), $1/\gamma_p$, becomes much shorter than the nonlinear eddy turnover time, $\tau_{nl}$ \cite{Loureiro2017,Mallet2017,Boldyrev2017,Comisso2018,Loureiro2020,note1}. Plasmoids develop in intense and elongated current sheets undergoing reconnection, and in MHD turbulence, the spatial scales and aspect ratios of those current sheets are controlled by the magnetic Reynolds number $R_m$, which quantifies the relative magnitudes of plasma convection and resistive diffusion. At sufficiently large $R_m$ ($> 10^5$), as in the solar atmosphere, magnetic reconnection is expected to be a ubiquitous phenomenon, as the thinning of current sheets can lead to a copious formation of plasmoids via the tearing instability \cite{Loureiro2007,Pucci2014,Comisso2016,UL16,Huang2017}, potentially changing the energy transfer across scales. 

Due to the complex, nonlinear nature of the turbulent energy cascade, direct numerical simulations (DNS) are likely the best means to investigate the role of magnetic reconnection in the energy transfer across scales and how it changes the turbulent energy spectra. To date, no evidence for a new range of the turbulent energy cascade due to magnetic reconnection has been provided by DNS in realistic three dimensions (3D) \cite{lazarian20203d,Schekochihin2022}. Such DNS are extremely challenging, mainly due to the high grid resolution required to capture the fine structure of the omnipresent current sheets in a turbulent plasma at large $R_m$. In addition, MHD turbulence and magnetic reconnection are known to behave differently in 2D and 3D \cite{Biskamp2003,lazarian20203d}, therefore 3D DNS with large $R_m$ are essential to fundamentally address this question.

\section*{RESULTS}
Here, we present the world's largest 3D MHD turbulence simulation (at a cost of $\sim 200$ million CPU hours) that self-consistently produces myriad fine current sheets. An elongated ($1 \times 1\times 2$) periodic box with $\sim 10000\times 10000 \times 5000$ grid cells was adopted to resolve the thin current sheets that develop at large magnetic Reynolds number, $R_m = 10^6$ (and the effective magnetic Reynolds number $R_{m,{\rm eff}} = 2\times10^5$ based on the energy injection scales). We initialized the simulations with uncorrelated, equipartitioned velocity and magnetic field fluctuations superimposed by a strong mean magnetic field in the elongated direction (see Materials and Methods for the detailed model setup). Such field configurations are common in a variety of astrophysical systems such as solar/stellar coronae and the interstellar medium. Compared with earlier 3D MHD turbulence simulations, the most prominent development here is that the current study reaches an unprecedented high-$R_m$ regime, such that the ubiquitous reconnecting current sheets in a turbulent bath become unstable to tearing instability.

The turbulent structures from our large-scale MHD simulation are visualized in Fig.~1. In Fig.~1a, volume rendering of the current density $|\mathbf{J}|$ depicts MHD turbulence in the entire simulation domain at the fully developed stage. Specifically, strong current sheets are ribbon-like (with large aspect ratios) and are aligned with the mean magnetic field $B_{z0}$ due to the parallel coherence of perturbations. Meanwhile, they are also current sheets in the perpendicular plane aligned with in-plane perturbed magnetic fields. These ribbon-shaped current sheets are subject to the tearing instability during their dynamical evolution \cite{Loureiro2007,Pucci2014,Comisso2016,UL16,Huang2017} (see Fig.~1d). The current density $|\mathbf{J}|$ in different x-y slices can be seen in movie~S1. One of these current sheets undergoing magnetic reconnection is highlighted in a zoomed-in subdomain (Figs.~1b and 1c), within which ``ripples'' induced by the formation of plasmoids/magnetic flux ropes are present on the current density isosurfaces. We cut a 2D slice across the embedded 3D magnetic flux ropes for $J_z$ (Fig.~1c; also see Fig.~5 in Materials and Methods for a selected flux-rope bundle at different viewing angles). The current sheet in the subdomain exhibits features similar to those observed in previous simulations of a single 3D reconnecting current layer \cite{Daughton2011,Huang2016,lazarian20203d}. This subdomain ($0.025\%$ of the complete volume) is representative of the types of coherent structures that develop throughout the entire domain. Reconnection-produced magnetic flux ropes in 3D exhibit a more complex morphology than their 2D counterpart. This is illustrated in Fig.~1d, in which the out-of-plane current density, $J_z$, in one arbitrary $x$-$y$ slice, points out that the reconnecting current sheet structures in 3D are fractured (or disrupted) and differ significantly from the island-like morphology in 2D \cite{Daughton2011,Huang2016,Dong2018,li2021three,pezzi2021current}, despite the similar size of the flux ropes.

Our simulation identifies a novel mechanism for energy transfer through the breakup of reconnecting current sheets into smaller fragments, which occurs generically in the large-$R_m$ regime studied here (Fig.~1d). The role of reconnection in controlling the energy transfer is supported by its imprint on the energy spectrum of the turbulent cascade in Fig.~2. At large scales, $k_\bot \lesssim k_*$, both the magnetic energy spectrum $E_B(k_\bot)$ and the kinetic energy spectrum $E_U(k_\bot)$ follow a power law with a slope of $-3/2$ (see dashed fitted lines), in agreement with expectations for the inertial range of a strong turbulent cascade mediated by Alfv\'{e}n waves when accounting for a reduction of nonlinearity due to dynamic alignment \cite{Boldyrev2006}. At scales $k_* \lesssim k_\bot \lesssim k_{\eta}$, current sheets in this tearing-mediated regime (i.e., reconnection-driven regime, as highlighted by the shaded region in Fig.~2) exhibit numerous reconnecting structures, or magnetic flux ropes, as shown in the inset. As a result, flux rope formation through reconnection dominates over MHD wave interactions in controlling turbulent energy transfer. Consequently, the energy spectrum in this range (termed as sub-inertial range hereafter) becomes steepened and is characterized by a spectral index of $-11/5$ when $R_m$ is sufficiently large, which is broadly consistent with the theoretical predictions \cite{Loureiro2017,Mallet2017,Boldyrev2017,Comisso2018}. It is noteworthy that the turbulent energy spectrum breaks at $k_* \approx 1000$ in Fig.~2, which also agrees with the theoretical prediction of tearing-disruption scale $k_* = R_{m,{\rm eff}}^{4/7} \approx 1000$ \cite{Loureiro2017,Mallet2017,Boldyrev2017,Comisso2018,note2}, where $R_{m,{\rm eff}} = 2\times10^5$ is the effective magnetic Reynolds number due to the energy injection scales as the energy spectrum peaks at $k_\bot \approx 30$. For comparison, we ran another 3D simulation with a relatively low magnetic Reynolds number, $R_m = 8 \times 10^4$, while keeping the initial condition identical. In this lower $R_m$ simulation, we observed that flux ropes are essentially absent (see Fig.~S1 in Supplementary Materials), which corroborates the lack of any imprint from reconnection-driven cascades in the corresponding energy spectrum (the blue curves in Fig.~2). We, therefore, conclude that a sufficiently large $R_{m}$ ($>10^5$) is one of the prerequisites for evolving current sheets to become tearing-unstable in a turbulent bath, consequently leading to the copious formation of flux ropes before reaching the scale given by $k_{\eta}$.

We further investigate changes in the shape (i.e., anisotropy) of statistical eddies in the sub-inertial range, one key element in making reconnection-driven energy cascade dominate over energy cascade through MHD wave interactions. In the top panels of Fig.~3, we report the dependence of the half width $\xi$ (panel a) and aspect ratio $\xi/\lambda$ (panel b) of the statistical eddies on the half thickness $\lambda$ based on second-order structure functions (see Fig.~6 in Materials and Methods). While the scaling $\xi/\lambda \propto \lambda^{-0.2}$ is observed in the traditional inertial range, in the sub-inertial range the scaling changes to approximately $\xi/\lambda \propto \lambda^{-0.1}$. The scaling is in contrast to the previous theoretical prediction $\xi/\lambda \propto \lambda^{4/5}$ \cite{Boldyrev2017}. Similarly, the bottom panels of Fig.~3 depict the relation between the half length $\ell_\parallel$ and the half thickness $\lambda$. A scaling relation $\ell_\parallel\propto\lambda^{1/2}$ is observed in the inertial range as expected from the theory. However, the scaling in the sub-inertial range, $\ell_\parallel\propto\lambda^{2/3}$ (and $\ell_\parallel/\lambda \propto \lambda^{-1/3}$), again deviates from the theoretical estimates \cite{Boldyrev2017}. It is noteworthy that the theoretical predictions $\xi/\lambda \propto \lambda^{4/5}$ and $\ell_\parallel/\lambda \propto \lambda^{1/5}$ in the sub-inertial range require the assumption $\theta_t \sim \delta/\zeta$, where $\theta_t$ is the alignment angle, and $\delta$ and $\zeta$ denote the inner layer width and the wavelength of the fastest growing tearing mode, respectively \cite{Boldyrev2017}. The difference between the theoretical predictions and the numerical experiments indicates that the assumption $\theta_t \sim \delta/\zeta$ needs further examination.

In order to address the underlying cause leading to this new range of the turbulent energy cascade, we also investigate the difference in the energy transfer between the inertial and sub-inertial ranges. For this purpose, we calculate the cylindrical shell-to-shell magnetic energy transfer function, $T_{bb}\left(Q,\ K\right)=-\int\mathbf{B}_{K}\cdot\left(\mathbf{u}\cdot\nabla\right)\mathbf{B}_{Q}{\rm d}^3x$, where $\mathbf{B}_{K}$ contains all Fourier modes $\tilde{\mathbf{B}}\left(\mathbf{k}\right)$ in the $K$-th perpendicular wavenumber shell, $k_{K}<k_{\perp}<k_{K+1}$, and $K$ is an integral shell number; $\mathbf{B}_{Q}\left(\mathbf{x}\right)$
is defined in the same way for the $Q$-th shell (see Materials and Methods for a detailed description). The red pixels in Fig.~\ref{fig:transfer-function}a represent positive magnetic energy transfer from the $Q$-th shell to the $K$-th shell, and the blue pixels the opposite. The diagonal pixels are white (i.e., empty) since there is no self-magnetic-energy-transfer by construction. The $T_{bb}$ distribution on the $k_{Q}$-$k_{K}$ phase plane in Fig.~\ref{fig:transfer-function}a is characterized by main features such as local, forward transfer. Figs.~\ref{fig:transfer-function}b and \ref{fig:transfer-function}c represent vertical cuts of $T_{bb}$ normalized to 
\begin{equation}
N_{{\rm inertial}}\sim\frac{E_{Q}}{\tau_{A}} (k_QL)^{-1/4}\sim\frac{E_{Q}^{3/2}k_{Q}}{\left(\rho V\right)^{1/2}}(k_QL)^{-1/4} \label{eq:Ninertial}
\end{equation}
and
\begin{equation}
N_{{\rm tearing}}\sim\gamma_p E_{Q}\sim\frac{\eta^{1/2}}{\left(\rho V\right)^{1/4}}E_{Q}^{5/4}k_{Q}^{3/2}, \label{eq:Nplasmoid}
\end{equation}
where $\tau_A$ is the Alfv\'{e}nic time scale and $\gamma_p$ the linear growth rate of the tearing instability (see Materials and Methods for details). The factor $(k_QL)^{-1/4}$ in Eq.~(\ref{eq:Ninertial}) takes into account the reduction of nonlinearity due to dynamic alignment \cite{Boldyrev2006}. Within the inertial range, energy transfer is expected to be self-similar, i.e., the energy transfer from the $Q$-th shell to the $K$-th shell operates in the same way as the transfer from the ($Q+1$)-th shell to the ($K+1$)-th shell. This is consistent with the observation in Fig.~\ref{fig:transfer-function}b that, when normalized to $N_{{\rm inertial}}$, the three cuts in the inertial range (solid lines in Fig.~\ref{fig:transfer-function}a) strongly overlap. In contrast, the three cuts in the tearing-mediated range (dotted curves in Fig.~\ref{fig:transfer-function}a) do not exhibit self-similarity in Fig.~\ref{fig:transfer-function}b, due to the underlying mechanism being different than the classical, inertial-range energy cascade. However, when $T_{bb}$ is normalized by $N_{{\rm tearing}}$, i.e., $T_{bb}/N_{{\rm tearing}}$, the cuts in the tearing-mediated range (dashed curve) become self-similar (Fig.~\ref{fig:transfer-function}c), thus confirming the existence of a sub-inertial range within which the energy transfer is controlled by the tearing instability in reconnecting current sheets.

\section*{DISCUSSION}

The present study suggests that the energy transfer in, e.g., the solar atmosphere at small scales can be fundamentally different from the classic paradigm of the turbulent energy cascade controlled by MHD wave interactions. Our calculated magnetic energy spectrum captures a new range of reconnection-driven energy cascade with a spectral index of $-2.2$. It is interesting to point out that the line-of-sight magnetic field observations of a coronal hole using magnetograms acquired by the Near InfraRed Imaging Spectrapolarimeter (NIRIS) operating at the Goode Solar Telescope of the Big Bear Solar Observatory (BBSO) showed that the turbulent magnetic energy spectra also exhibit a new range with a spectral index of -2.2 \cite{Abramenko2020} (also see Fig.~S2 in Supplementary Materials), but it requires further observations to rule out the coincidence and then to identify the underlying mechanism that leads to the spectral break. On the other hand, similar observations acquired by the Helioseismic and Magnetic Imager (HMI) onboard the Solar Dynamic Observatory (SDO) with a relatively low spatial resolution did not present a steepening in the corresponding energy spectra, suggesting that a high spatial resolution might be required to reveal this new range of energy transfer in the turbulent solar atmosphere. To address the discrepancy between the observations of NIRIS and HMI, we illustrated, with the same dataset in Fig.~S2, that the spatial resolution of observation can significantly influence the scientific findings. 

In addition, spacecraft such as Solar Orbiter may also be able to reveal this newly identified sub-inertial range in magnetic energy spectra through remote sensing observations. Recently, Extreme Ultraviolet Imager (EUI) onboard Solar Orbiter observed transient small-scale brightenings prevalent in the corona of the quiet Sun termed `campfires' \cite{Berghmans2021}. It has been proposed that the majority of campfire events observed by EUI are driven by magnetic reconnection, which may play an important role in the coronal heating of the quiet Sun \cite{Chen2021,Clery2021}. Our 3D simulation results in the large-$R_m$ regime suggest that magnetic reconnection is a ubiquitous process in the turbulent solar corona where the $R_m$ is even larger and thus the current sheets can thin down to much smaller scales and form the fractal structures within which copious formation of plasmoids occurs. Hence, there should be many more reconnection sites than observed, which can be revealed by (future) high-resolution extreme ultraviolet images.  
 
In addition to the solar coronal heating \cite{Chen2021,Berghmans2021}, the omnipresence of plasmoids associated with magnetic reconnection, identified in this study, also has broad implications and consequences on particle acceleration in the solar/stellar coronae, accretion disks, and jets from compact objects \cite{Drake06,Comisso2022}, and on filament formation in the \emph{Herschel} maps of the Orion A giant molecular cloud \cite{Kong2021ApJ}. Moreover, the ubiquitous presence of dense plasmoids in the interstellar medium can also play an important role in pulsar scintillation \cite{Narayan92}. Indeed, all the aforementioned astrophysical systems are associated with large magnetic Reynolds numbers ($R_m > 10^5$), and thus could be characterized by the tearing-mediated turbulence identified in the present work.

\begin{figure*}
\begin{center}
\includegraphics[width=1.0\columnwidth]{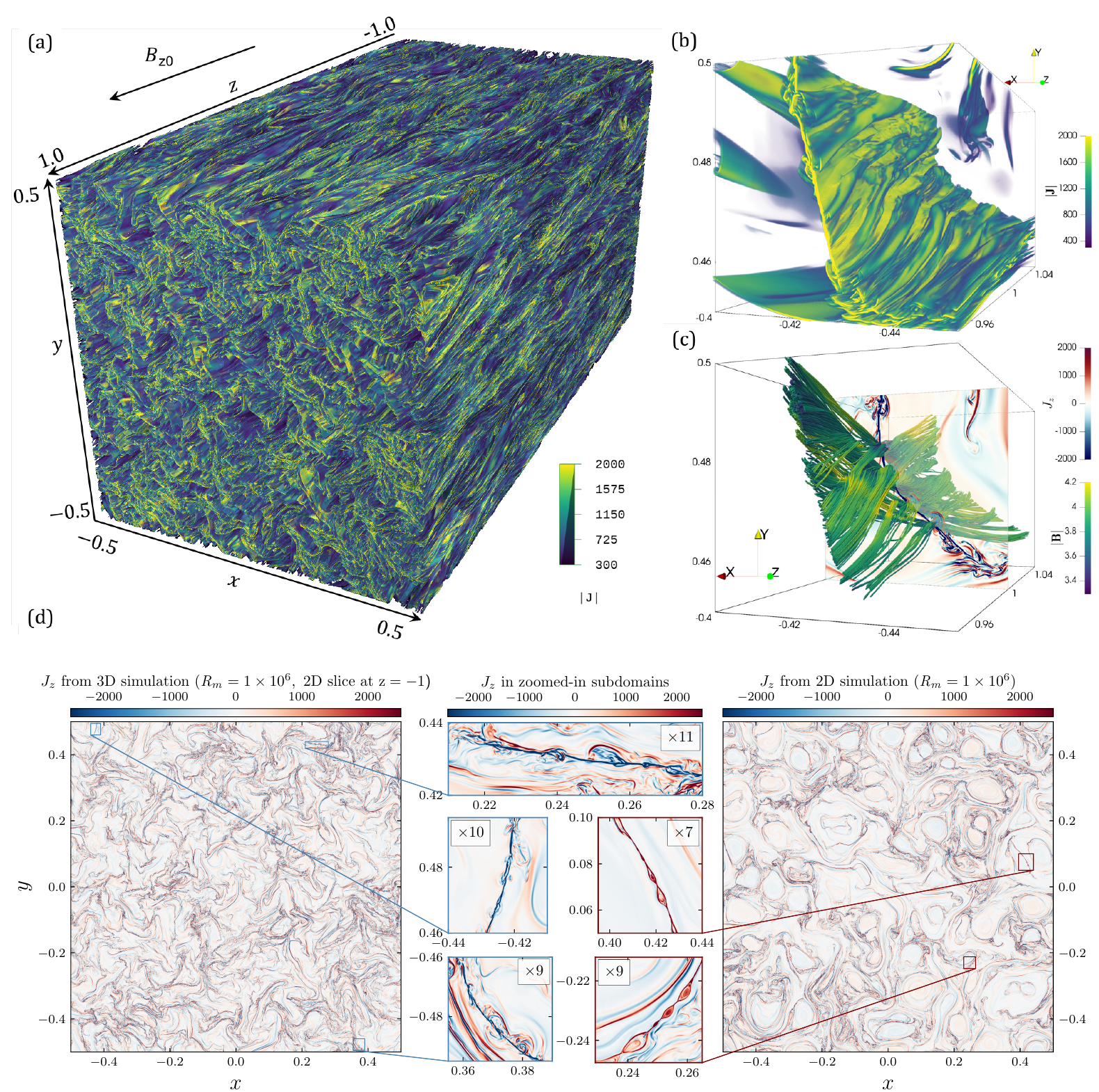}
\end{center}
\end{figure*}

\begin{figure*}
\caption{{\bf Reconnecting current sheets and magnetic flux ropes in MHD turbulence.} ({\bf a}) Volume rendering of the current density $|\mathbf{J}|$ in the entire domain at a stage when turbulence is fully developed. Myriad of current sheets is evident in the plane perpendicular to the mean magnetic field $B_{z0}$. Panels ({\bf b}) and ({\bf c}) depict one reconnecting current sheet and the embedded flux-ropes in a small subdomain (within the boundaries $[-0.45, -0.4]\times[0.45, 0.5]\times[0.95, 1.05]$). Panel ({\bf b}) shows the volume rendering of $|\mathbf{J}|$, while panel ({\bf c}) displays magnetic field lines (colored by $|\mathbf{B}|$) associated with the featured current sheet (including magnetic flux ropes) and a $x$-$y$ slice view of the current density component, $J_z$, along the mean magnetic field. ({\bf d}) Out-of-plane current density $J_z$ in a $x$-$y$ slice (at $z=-1$) of the 3D turbulence simulation (left) compared with the corresponding result from a 2D simulation (right). Copious formation of magnetic flux ropes/plasmoids occurs in both 3D and 2D simulations despite the different morphology. Zoomed-in subdomains are used to illustrate the increased morphological complexity that characterizes the 3D simulation.}
\end{figure*}
\newpage

\begin{figure*}
\begin{center}
\includegraphics[width=\columnwidth]{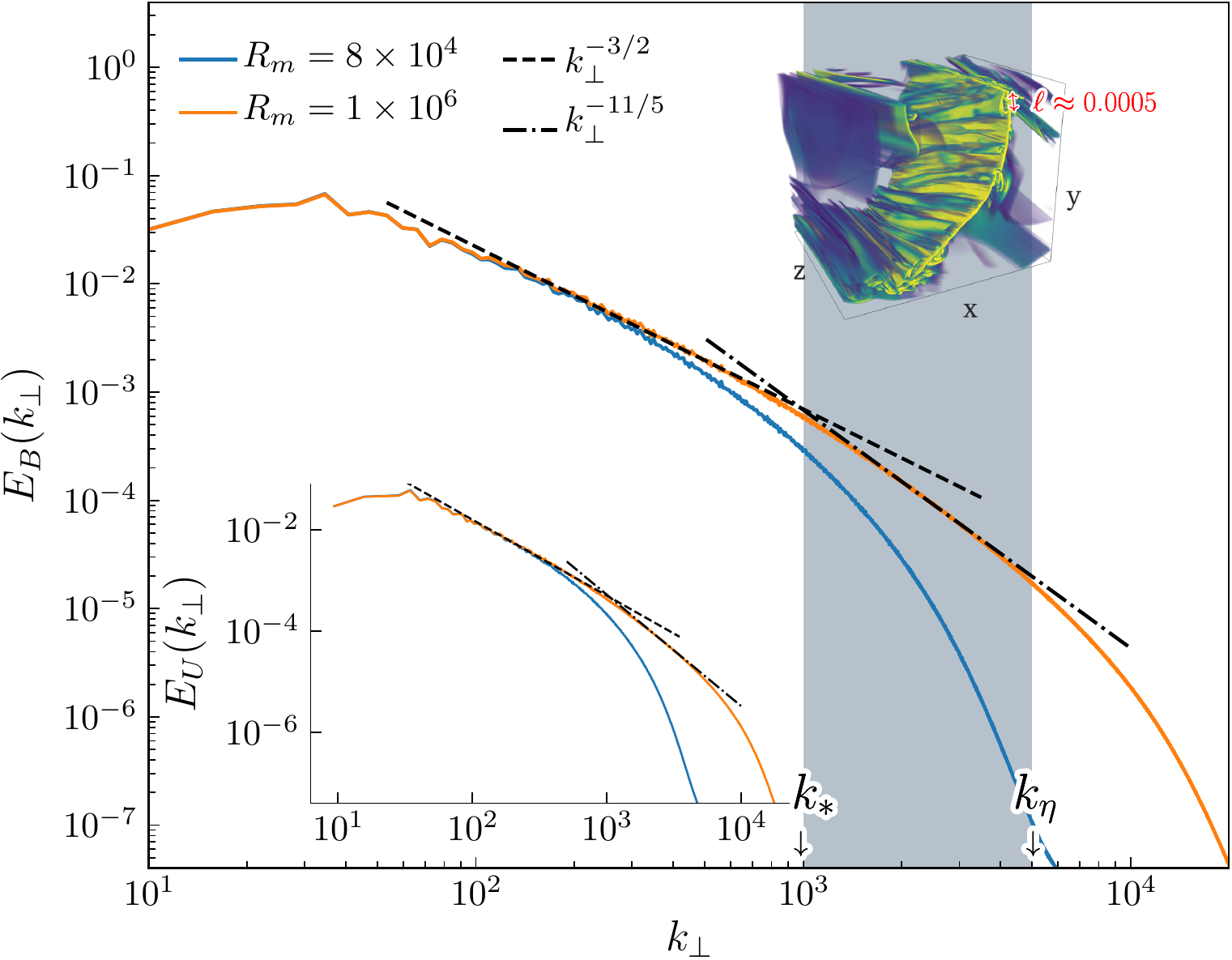}
\end{center}
\end{figure*}

\begin{figure*}
\caption{{\bf Steepening of energy spectra in reconnection-driven energy cascade.} Field-perpendicular magnetic and kinetic energy spectra, $E_B(k_\bot)$ and $E_U(k_\bot)$, showing a standard inertial range with a slope $E_{B,U}(k_\bot) \propto k_\bot^{-3/2}$ and a reconnection-driven (or tearing-mediated) sub-inertial range with a slope $E_{B,U}(k_\bot) \propto k_\bot^{-11/5}$ for large magnetic Reynolds number $R_m$ (orange curves). The sub-inertial range is absent in an equivalent 3D simulation with lower $R_m$ (blue curves). The shaded area emphasizes the reconnection-driven sub-inertial range, with wavenumbers corresponding to the typical transverse scales of the flux ropes. The left edge of the shaded range is identified at $k_* \approx 10^3\approx R_{m,{\rm eff}}^{7/4}$, where $R_{m,{\rm eff}}\approx 2\times10^5$ is computed with the injection scale at $k \approx 30$. The right edge, marked by $k_{\eta}$
, is the dissipation scale defined such that $\eta \int_{0}^{k_{\eta}} E_B(k_\bot) k_\bot^2 {\rm d}k_\bot$ accounts for approximately half of the resistive dissipation power $\eta \langle J^2 \rangle$.
The small volume rendering inset illustrates a typical reconnecting current layer, that is ubiquitous in this range, with the transverse size of an embedded flux rope annotated in red.
}
\end{figure*}
\newpage

\begin{figure*}
\includegraphics[width=1\columnwidth]{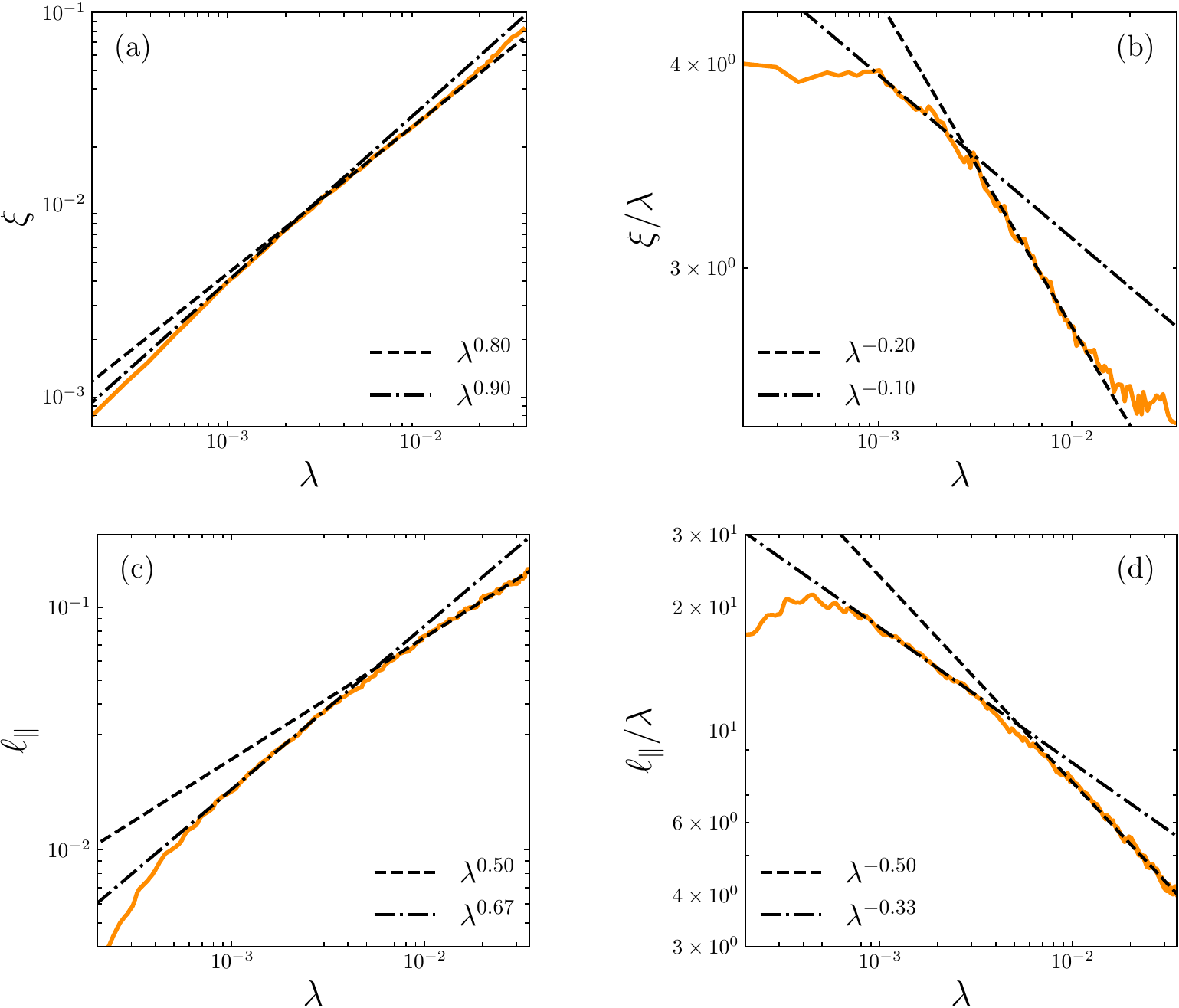}
\caption{\label{fig:main-structure-function}{\bf Scale-dependent anisotropy of statistical eddies.} ({\bf a}) The perpendicular half width $\xi$ of the statistical eddies as a function of the perpendicular half thickness $\lambda$. This function is obtained by equating the second-order $\mathbf{B}$-trace structure function measured along $\xi=0$ and $\lambda=0$ axes in Fig.~\ref{fig:method-structure-function}b. ({\bf b}) Aspect ratio $\xi/\lambda$ of the statistical eddies as a function of $\lambda$. (\textbf{c}) and (\textbf{d}) are the scaling relation between the parallel half length $\ell_{\parallel}$ and $\lambda$, obtained in a similar way but in the local coordinate plane $\xi=0$. The fitted dashed and dot-dashed lines show two scaling laws in the traditional inertial range and the tearing-mediated sub-inertial range of the turbulent energy cascade.}
\end{figure*}
\newpage

\begin{center}
\begin{figure*}
\includegraphics[width=1\columnwidth]{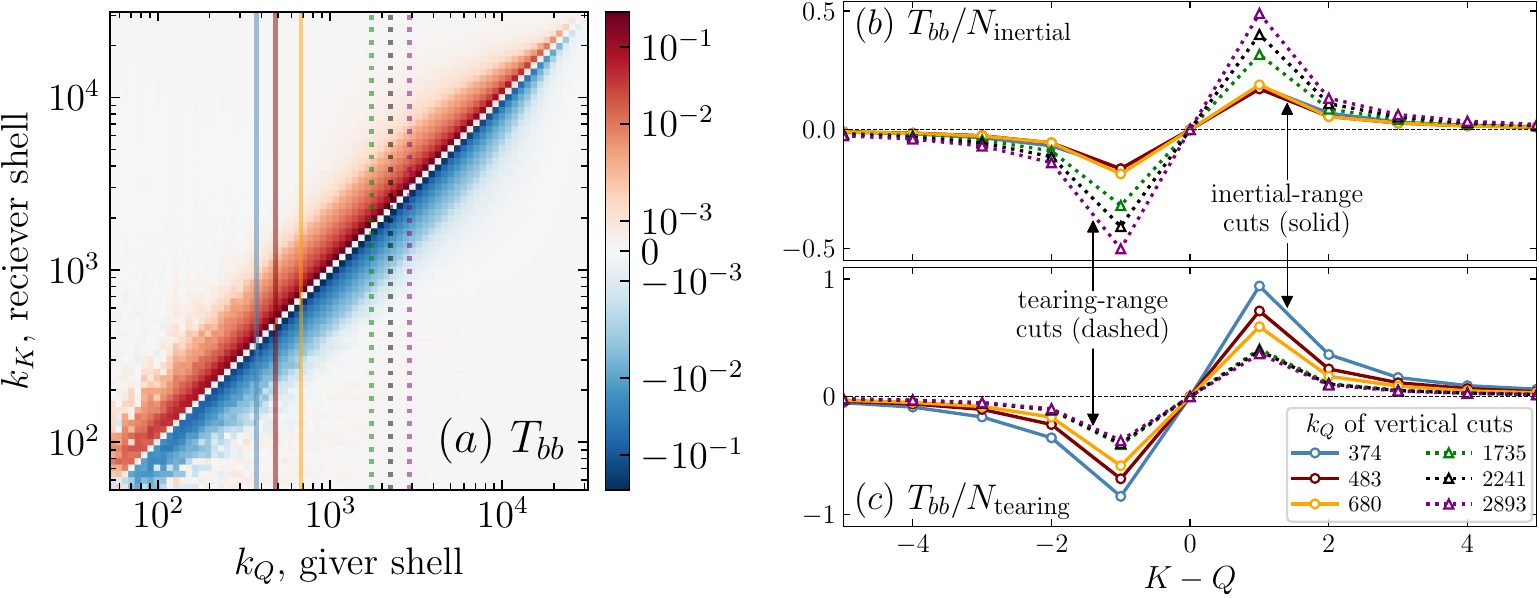}
\caption{\label{fig:transfer-function}\textbf{Self-similarity of normalized cylindrical shell-to-shell energy transfers.} \textbf{(a)} Magnetic-to-magnetic energy transfer
functions $T_{bb}\left(Q,K\right)=-\int\mathbf{B}_{K}\cdot\left(\mathbf{u}\cdot\nabla\right)\mathbf{B}_{Q}{\rm d}^3{x}$ in the $k_{Q}$-$k_{K}$ plane. \textbf{(b)-(c)} Normalized energy transfer function values along different vertical cuts in specific giver $k_{Q}$ shells. The cut locations are marked in \textbf{(a)}, with solid-line cuts chosen in the inertial range and dotted-line cuts in the reconnection-driven (or tearing-mediated) sub-inertial range. The normalization used in \textbf{(b)} and \textbf{(c)}, $\ensuremath{N_{{\rm inertial}}=E_{Q}^{3/2}k_{Q}}(k_QL)^{-1/4}/(\rho V)^{1/2}$ and $N_{{\rm tearing}}=\eta^{1/2}E_{Q}^{5/4}k_{Q}^{3/2}/(\rho V)^{1/4}$ (see Materials and Methods), are appropriate for the inertial range and the tearing-mediated sub-inertial range, respectively.
The horizontal coordinates in \textbf{(b)} and \textbf{(c)} are the shell number differences
(integers), $K-Q$.}
\end{figure*}
\par\end{center}
\newpage

\section*{MATERIALS AND METHODS}

\subsection*{Model description}

Following procedures previously described in \cite{Dong2018}, the governing equations of our numerical model are the dimensionless visco-resistive MHD equations:
\begin{equation} \label{continuity_eq}
\partial_t \rho  + \nabla \cdot (\rho {\bf{u}}) = 0 \, ,
\end{equation}
\begin{equation} \label{}
\partial_t (\rho {\bf{u}}) + \nabla  \cdot (\rho {\bf{uu}}) =  - \nabla \left( p + B^2/2 \right) + \nabla  \cdot ({\bf{BB}}) + \nu {\nabla ^2}(\rho {\bf{u}}) \, ,
\end{equation}
\begin{equation} \label{}
\partial_t p + \nabla  \cdot (p {\bf{u}}) = (\gamma - 1) \left( { - p \nabla \cdot {\bf{u}} + \eta {\bf{J}}^2} \right) \, ,
\end{equation}
\begin{equation} \label{induction_eq}
\partial_t {\bf{B}} = \nabla  \times ({\bf{u}} \times {\bf{B}} - \eta {\bf{J}}) \, ,
\end{equation}
where $\rho$, ${\bf{u}}$ and $p$ are the mass density, velocity, and pressure of the plasma, respectively; ${\bf{B}}$ is the magnetic field and ${\bf{J}} = \nabla  \times {\bf{B}}$ denotes the electric current density. The kinematic viscosity and the magnetic diffusivity are denoted  as $\nu$ and $\eta$, respectively, while $\gamma$ is the adiabatic index.

\subsection*{Model setup}\label{subsec:ModelSetup}

We solve Eqs.~(\ref{continuity_eq})-(\ref{induction_eq}) using the BATS-R-US MHD code \cite{Toth2012} by adopting a fifth-order scheme \cite{Chen2016} in a domain $\{(x,y,z):-L_0/2 \leq x, y \leq L_0/2, -L_0 \leq z \leq L_0\}$, where $L_0$ is set to unity. Periodic boundary conditions are employed in all three directions. Lengths are normalized to the box size $L_0$, velocities to the characteristic Alfv\'{e}n speed $V_A$, and time to $L_0/V_A$. We initialize the simulations by placing uncorrelated, equipartitioned velocity and magnetic field fluctuations in Fourier harmonics as follows,
\begin{eqnarray}
\psi &=& \sum \limits_{l,m,n}(a_{mn}/2\pi)\sin(2\pi mx/L_x + 2\pi ny/L_y + 2\pi lz/L_z + \phi_{lmn}) \\
B_x &=& \frac{\partial \psi}{\partial y} = \sum \limits_{l,m,n} a_{mn} (n/L_y) \cos(2 \pi mx/L_x + 2 \pi ny/L_y + 2 \pi lz/L_z + \phi_{lmn}) \\
B_y &=& -\frac{\partial \psi}{\partial x} = - \sum \limits_{l,m,n} a_{mn} (m/L_x) \cos(2 \pi mx/L_x + 2 \pi ny/L_y + 2 \pi lz/L_z + \phi_{lmn}) \\
B_z &=& B_{z0} 
\end{eqnarray}
where we set 
\begin{equation}
a_{mn} = \frac{\sqrt{2}B_0}{N^{1/2}(m^2/L_x^2+n^2/L_y^2)^{1/2}}
\end{equation}
and $L_x = L_y = L_0$, and $L_z = 2 L_0$. $\phi_{lmn}$ denotes random phases for each mode. The summation is over the range $0 \le m \le m_{max}$, $-n_{max} \le n \le n_{max}$. However, for $m=0$, we limit to the range to $0<n<n_{max}$. The range of $l$ is always $-l_{max} \le l \le l_{max}$. The total number of modes are 
\begin{equation}
N = (2l_{max}+1)[(2m_{max}+1)(2n_{max}+1)-1]/2 .
\end{equation}
We only sum over half of the Fourier space because the $(m,n)$ mode and the $(-m,-n)$ mode are not independent. Here, we choose $m_{max} = n_{max} = 10$ and $l_{max} = 5$. The above setup leads to 
\begin{equation}
\langle B_{\perp}^2 \rangle= B_0^2 ,
\end{equation}
where $\langle ... \rangle$ represents the spatial average. 

For the velocity $\mathbf{u}$, we use the similar expressions:
\begin{eqnarray}
u_x &=& \sum \limits_{l,m,n} b_{mn} (n/L_y) \cos(2 \pi mx/L_x + 2 \pi ny/L_y + 2 \pi lz/L_z + \phi^{\prime}_{lmn}) \\
u_y &=& - \sum \limits_{l,m,n} b_{mn} (m/L_x) \cos(2 \pi mx/L_x + 2 \pi ny/L_y + 2 \pi lz/L_z + \phi^{\prime}_{lmn}) \\
u_z &=& 0 
\end{eqnarray}
where we set 
\begin{equation}
b_{mn} = \frac{\sqrt{2}u_0}{N^{1/2}(m^2/L_x^2+n^2/L_y^2)^{1/2}} .
\end{equation} 
Similarly, we have
\begin{equation}
\langle u_{\perp}^2 \rangle= u_0^2 .
\end{equation}

\begin{figure*}
\begin{centering}
{\includegraphics[width=1.0\columnwidth]{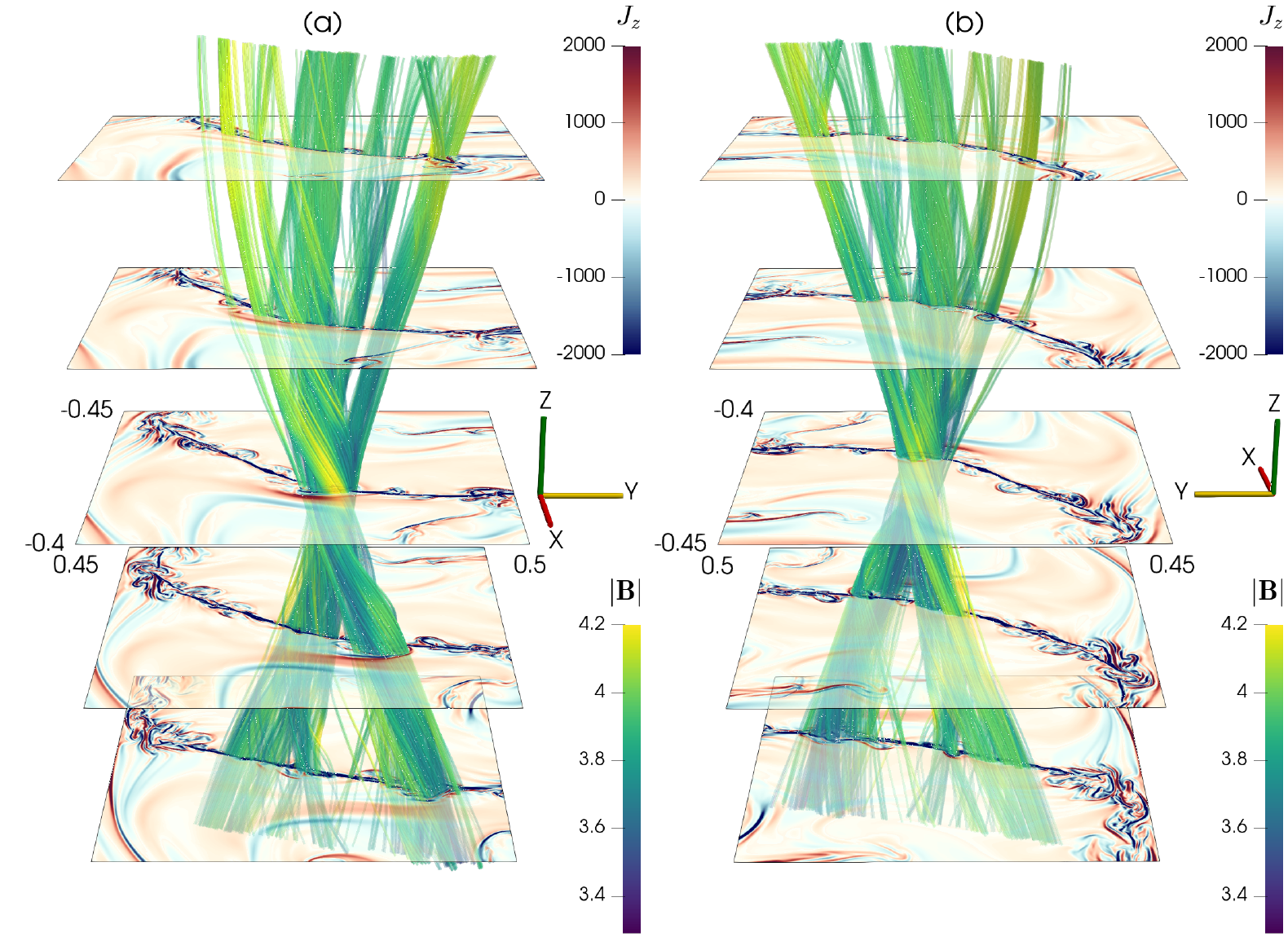}}
\caption{\label{fig:method-3D-flux-ropes}{\bf 3D magnetic flux ropes with elongated current sheets.} Flux ropes passing through a plasmoid located near the center of the current sheet displayed in the intermediate 2D slice. The chosen views are along the $+x$ (panel {\bf a}) and $-x$ (panel {\bf b}) directions, respectively. Magnetic field lines are colored accordingly to the magnetic field magnitude $|\mathbf{B}|$, while the different $x$-$y$ slices are colored by the value of the electric current density component, $J_z$, in the direction of the mean magnetic field.}
\par\end{centering}
\end{figure*}

The plasma density and pressure are initially set to constant values $\rho = 1.0$ and $p = 1.6$, respectively. The constants $B_0$ and $u_0$ determine the strength of the initial velocity and magnetic field fluctuations. In this work, we set $B_0 = u_0 = 1$, which gives the initial turbulent energy $E = \frac{1}{2}\langle u_{\perp}^2 +  B_{\perp}^2 \rangle = \frac{1}{2}(u_0^2+B_0^2) = 1$. The mean magnetic field $B_{z0} = 4$ leads to a plasma beta (ratio of the plasma pressure to the magnetic pressure) $\beta \approx 0.2$. In the main 3D simulation presented here, the magnetic Reynolds number is $R_m = u_0L_0/\eta = 10^6$, where $\eta = 10^{-6}$ denotes the magnetic diffusivity. The effective magnetic Reynolds number is $R_{m,{\rm eff}} = 2\times10^5$ due to the energy injection scales as the energy spectrum peaks at $k_\bot \approx 30$ (see Fig.~2). We set the viscosity $\nu = 10^{-6}$, such that the magnetic Prandtl number $P_m = \nu/\eta = 1$. This large $R_m$ value allows the copious formation of flux ropes within the current sheets, as verified by visual inspection of the simulation data in a tiny subdomain (see example in Fig.~\ref{fig:method-3D-flux-ropes}). For the 2D simulations, we adopt similar initial setups but ignore the third dimension in $z$. For both 3D and 2D studies, we also ran one additional simulation with $R_m = 8 \times 10^4$ and a period box of $\sim 2000\times 2000 \times 1000$ and $\sim 2000\times 2000$ grid cells, respectively (see Fig.~S1 in Supplementary Materials). Compared with the main 3D simulation (at a cost of about 200 million CPU hours with approximately 0.5 trillion grid cells), the computational cost of the rest simulations is negligible.

\subsection*{Second-order structure function} 
\begin{figure*}
\begin{centering} 
\includegraphics[width=1.0\columnwidth]{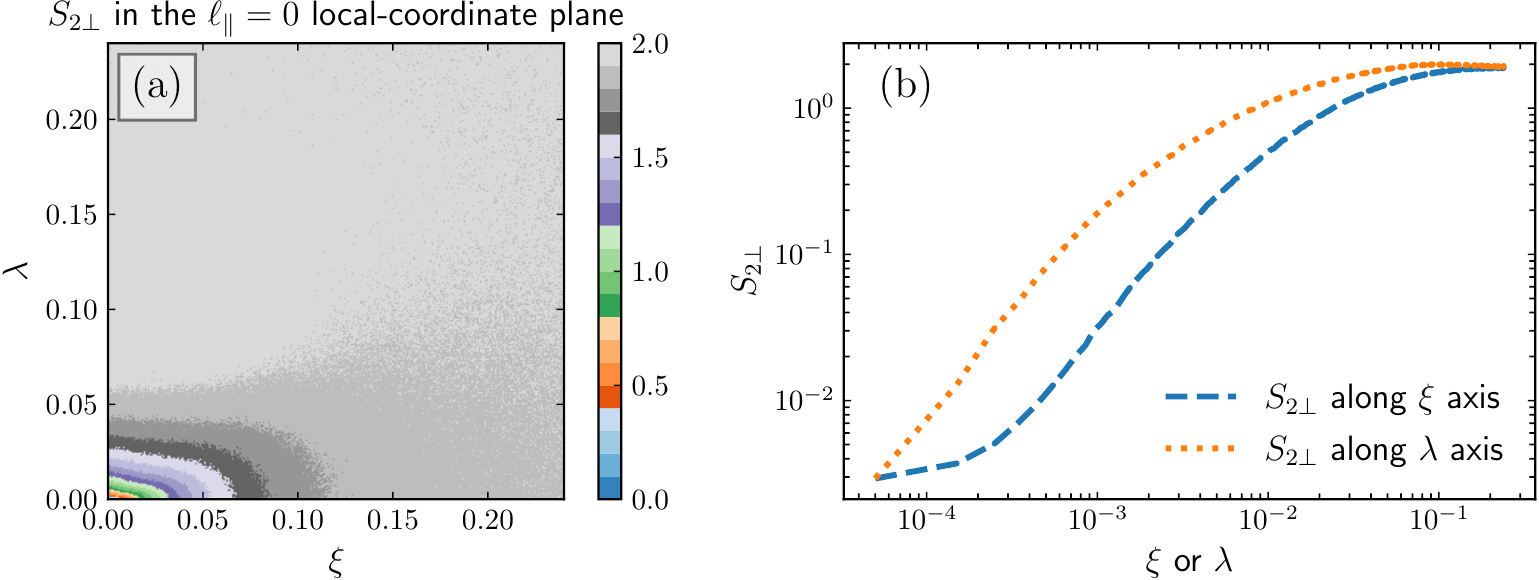}
\caption{\label{fig:method-structure-function}{\bf Anisotropy of the statistical eddies perpendicular to the local mean magnetic field.} 
({\bf a}) Cross sections of the statistical eddies at different values from the second-order structure function $S_{2\perp}=\left< \left| \mathbf{B}_1-\mathbf{B}_2 \right|^2\right>$. 
({\bf b}) $S_{2\perp}$ as a function of the perpendicular half width $\xi$ or the perpendicular half thickness $\lambda$ of the statistical eddies.}
\par\end{centering}
\end{figure*}

The anisotropy of statistical eddies shown in Fig.~\ref{fig:main-structure-function} is calculated based on a second-order $\mathbf{B}$-trace structure function. To calculate
this structure function, we randomly sample a large number of data-point pairs. For each pair of points, 1 and 2, the displacement $\delta\mathbf{r}=\mathbf{r}_{1}-\mathbf{r}_{2}$ is projected onto a local, scale-dependent coordinate system $\left(\ell_{\parallel},\ \xi,\ \lambda\right)$: $\hat{\ell}_{\parallel}$
is along the local mean field $\mathbf{B}_{{\rm local}}=\left(\mathbf{B}_{1}+\mathbf{B}_{2}\right)/2$,
$\hat{\xi}$ is along the local perpendicular fluctuation field $\delta\mathbf{B}_{\perp,{\rm local}}=\mathbf{B}_{{\rm local}}\times\left[\left(\mathbf{B}_{1}-\mathbf{B}_{2}\right)\times\mathbf{B}_{{\rm local}}\right]/|\mathbf{B}_{{\rm local}}|^2$, and $\hat{\lambda}$ completes the right-handed coordinate system. We then accumulate the contributions from all sampling pairs to compute the $\mathbf{B}$-trace structure function $S_{2}=\left\langle \left|\mathbf{B}_{1}-\mathbf{B}_{2}\right|^{2}\right\rangle$, an ensemble-averaged function of $\left(\ell_{\parallel},\ \xi,\ \lambda\right)$. As an example, here we show the cross-section of $S_{2}$ in the $\ell_{\parallel}=0$ plane. The consequent perpendicular structure function $S_{2\perp}\left(\xi,\ \lambda\right)$ is illustrated in Fig.~\ref{fig:method-structure-function}a, which exhibits clear anisotropy between the perpendicular half width, $\xi$, of the statistical eddies and the perpendicular half thickness, $\lambda$. It is immediately observed that Fig.~\ref{fig:method-structure-function}a also depicts the perpendicular cross-section of statistical eddies at different scales.

To further quantify the anisotropy, we investigate the aspect ratio of the perpendicular eddies, $\xi/\lambda$. To this end, we measure $S_{2\perp}$ along the two perpendicular axes, $S_{2\perp}\left(\xi;\ \lambda=0\right)$ and $S_{2\perp}\left(\lambda;\ \xi=0\right)$, as shown in Fig.~\ref{fig:method-structure-function}b, and then find a mapping between $\xi$ and $\lambda$ by equating $S_{2\perp}$ along the two curves. This gives the result illustrated in Fig.~\ref{fig:main-structure-function}. 

Following a similar procedure, we also obtain the relation between the parallel half length, $\ell_{\parallel}$, and the perpendicular half thickness, $\lambda$, in Fig.~\ref{fig:main-structure-function} based on the structure function $S_{2}$ in the $\xi=0$ plane.

\subsection*{Shell-to-shell energy transfers in the inertial and tearing-mediated sub-inertial ranges}

To further investigate the role of tearing instability in the turbulent energy cascade, we calculated the cylindrical shell-to-shell magnetic energy transfer function \cite{Alexakis2007} 
\begin{equation}
T_{bb}\left(Q,\ K\right)=-\int\mathbf{B}_{K}\cdot\left(\mathbf{u}\cdot\nabla\right)\mathbf{B}_{Q}{\rm d}^3x.\label{eq:Tbb}
\end{equation}
Here, 
\begin{equation}
\mathbf{B}_{K}\left(\mathbf{x}\right)=\int_{k_{K}<k_{\perp}<k_{K+1}}\tilde{\mathbf{B}}\left(\mathbf{k}\right)e^{i\mathbf{k}\cdot\mathbf{x}}{\rm d}^3k\label{eq:Bk}
\end{equation}
contains all Fourier modes $\tilde{\mathbf{B}}\left(\mathbf{k}\right)$
in the $K$-th perpendicular wavenumber shell, $k_{K}<k_{\perp}<k_{K+1}$,
and $K$ is an integral shell number. $\mathbf{B}_{Q}\left(\mathbf{x}\right)$
is defined in the same way for the $Q$-th shell. $T_{bb}\left(Q,K\right)$
then gives the transfer rate of magnetic energy from the $Q$-th shell to
the $K$-th shell \cite{Alexakis2005,Debliquy2005}, or, correspondingly, from the spatial scale $\sim1/k_{Q}$ to the scale $\sim1/k_{K}$.
We adopted 100 logarithmic bins along the $K$ and $Q$ directions
between $k_{\perp}=2\pi$ and $k_{\perp}=10^{4}\pi$. The resulting
$T_{bb}$ values from our simulation on the $k_{Q}$-$k_{K}$ plane
is shown in Fig.~\ref{fig:transfer-function}a. The dominance of red pixels above
the diagonal where $k_{K}>k_{Q}$ confirms the forward transfer from larger to smaller scales. 
Fig.~\ref{fig:transfer-function}a also shows that transfers occur primarily close to the diagonal, indicating that the energy transfer is mostly local.

\subsection*{Normalization of the cylindrical shell-to-shell energy transfer rate} 

In the inertial range, we assume that Boldyrev's turbulence theory with dynamic alignment holds \cite{Boldyrev2006}. In this picture, turbulence eddies are anisotropic in all three directions, with dimensions $\xi$ and $\lambda$ perpendicular to the local field, and $\ell_\parallel$ along the local field. These scales are related as $\xi \sim L(\lambda/L)^{3/4}$ and $\ell_\parallel \sim L(\lambda/L)^{1/2}$, where $L$ is the large-scale length. The magnetic fluctuation $B_{\lambda} \sim B_0(\lambda/L)^{1/4}$ at scale $\lambda$, where $B_0$ is the large-scale magnetic field. The energy cascade of eddies at scale $\lambda$ occurs on the time scale $\tau_\lambda \sim \ell_\parallel/V_{A0} \sim \xi/V_{A\lambda}$. Let $E_Q$ be the magnetic energy in the $Q$-th shell (i.e., the perpendicular wavenumber $k_\bot$ satisfies $k_{Q}<k_{\perp}<k_{Q+1}$) and $V$ be the volume of the domain. The magnetic energy density $E_Q/V$ on the $\lambda_Q\sim 1/k_Q$ scale  is proportional to $B_Q^2/2$; therefore, the Alfv\'{e}n speed $V_{AQ}= B_Q/\sqrt{\rho} \sim E_Q^{1/2}/(V\rho)^{1/2}$ and the cascade time $\tau_Q \sim \xi_Q/V_{AQ} \sim (k_QL)^{1/4}(V\rho)^{1/2}/E_Q^{1/2}k_Q$. This leads to the normalization $N_{{\rm inertial}}\sim{E_{Q}}/{\tau_{Q}}\sim (k_QL)^{-1/4}{E_{Q}^{3/2}k_{Q}}/{\left(\rho V\right)^{1/2}}$. We expect $T_{bb}\left(Q,K\right)/N_{{\rm inertial}}$ in Fig.~4b to be approximately
independent of $Q$ in the inertial range. 

In the tearing-mediated range, the energy cascade time is governed by $1/\gamma_p$,
where $\gamma_p$ is the linear growth rate of the tearing instability. For a current sheet with a half thickness $\lambda$ and an upstream Alfv\'en speed $V_{A\lambda}$, the linear growth rate \cite{Biskamp2000} $\gamma_p \sim (V_{A\lambda}\lambda/\eta)^{-1/2}V_{A\lambda}/\lambda$. For the $Q$-th shell, $\lambda \sim 1/k_Q$, $V_{A\lambda} \sim E_Q^{1/2}/(V\rho)^{1/2}$. Hence, we propose another normalization
for the reconnection-driven (or tearing-mediated) sub-inertial range, 
$N_{{\rm tearing}}\sim \gamma_p E_{Q}\sim {\eta^{1/2} E_{Q}^{5/4}k_{Q}^{3/2}/\left(\rho V\right)^{1/4}}$. We expect $T_{bb}\left(Q,K\right)/N_{{\rm tearing}}$ in Fig.~4c to be nearly independent of $Q$ in the tearing-mediated sub-inertial range. 

In our simulation, the plasma density $\rho \simeq 1$ and the volume $V=2$. We adopt the value $\rho V=2$ in the normalization for producing Fig.~\ref{fig:transfer-function}.

\newpage
\begin{centering}
\textbf{\Large Supplementary Materials for \\ Reconnection-Driven Energy Cascade in Magnetohydrodynamic Turbulence}
\end{centering}

\textbf{This PDF file includes:}\\
Fig. S1. Current sheets in low-$\boldsymbol{R_m}$ 3D versus 2D MHD turbulence. \\
Fig. S2. Cell-averaged simulation results and comparison to spectral observations. \\
Section S1. Comparative simulations with a low magnetic Reynolds number. \\
Section S2. Exploring the connections to solar observations. \\
Movie S1. Animation of current density $|J|$. \\

\section*{Section S1. Comparative simulations with a low magnetic Reynolds number.}
For both 3D and 2D studies, we ran additional simulations with relatively low magnetic Reynolds number, $R_m=8 \times 10^4$ (or $R_{m,{\rm eff}} = 1.6\times10^4$), at resolutions of $\sim 2000\times 2000 \times 1000$ and $2000\times2000$ grid cells, respectively. We set the magnetic diffusivity $\eta = 1.25\times10^{-5}$ and viscosity $\nu = 1.25\times10^{-5}$, such that the magnetic Prandtl number $P_m = \nu/\eta = 1$. These two additional simulations are used to comparatively demonstrate that only for large $R_m$ should magnetohydrodynamic turbulence be able to show the copious formation of magnetic flux ropes (or plasmoids). Fig.~S1 depicts the electric current density component, $J_z$, in a 2D cut (at $z=-1$) of the low-$R_m$ 3D simulation and in the low-$R_m$ 2D simulation. Different from Fig.~1d in the main narrative, these low-$R_m$ simulations do not produce a significant amount of magnetic flux ropes. This is because, for the low $R_m$, the evolving current sheets cannot reach sufficiently fine scales to trigger the tearing instability \cite{Loureiro2007,Pucci2014,Comisso2016,UL16,Huang2017} before being dissipated at scales $\sim 1/k_{\eta}$.

\begin{figure*}
\begin{centering}
\includegraphics[width=1\columnwidth]{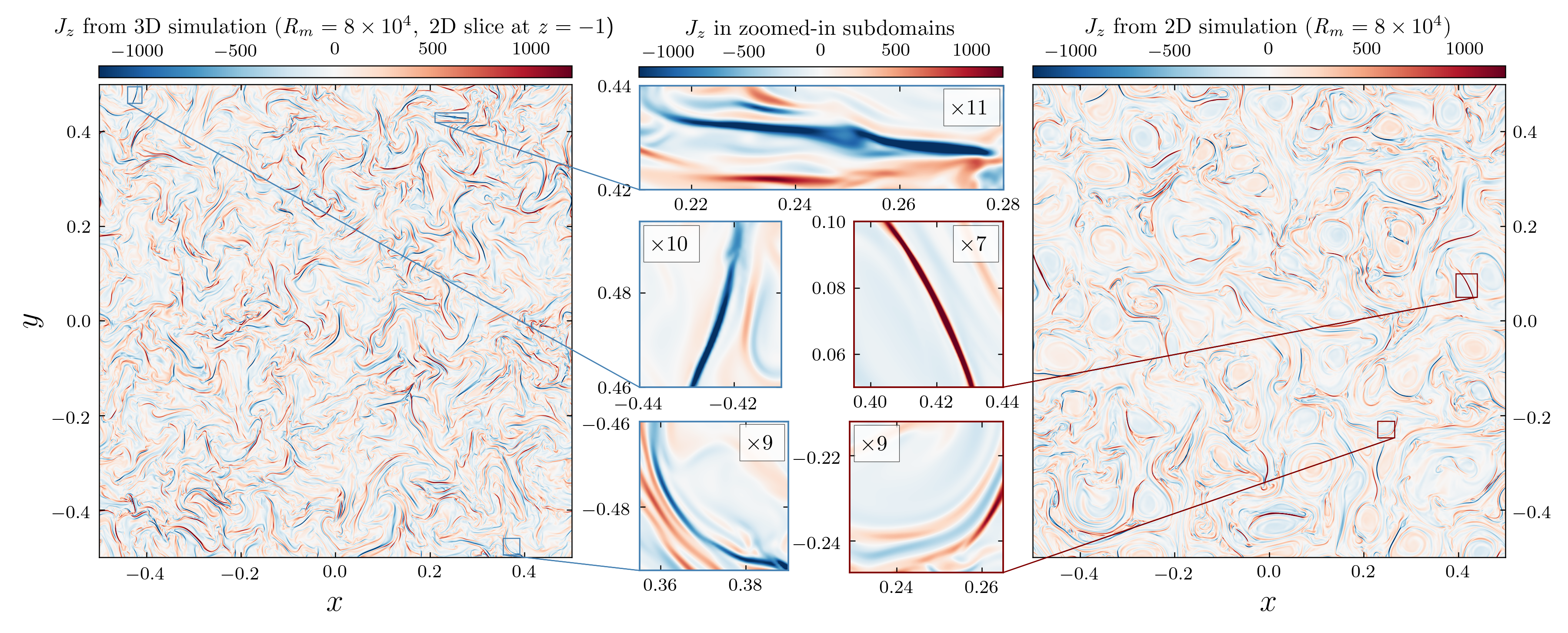}
\caption{{\bf Current sheets in low-$\boldsymbol{R_m}$ 3D versus 2D MHD turbulence.} This figure (with low-$R_m$) is analogous to Fig.~1d in the main narrative from large-$R_m$ simulations. Out-of-plane current density $J_z$ in a $x$-$y$ slice (at $z=-1$) of the 3D low-$R_m$ turbulence simulation (left) compared with the corresponding result from the 2D low-$R_m$ simulation (right). The zoomed-in subdomains coincide with those in Fig.~1d. Compared to Fig.~1d, it is clear that these low-$R_m$ simulations do not produce the copious formation of plasmoids.}
\par\end{centering}
\end{figure*}

\section*{Section S2. Exploring the connections to solar observations.}

We mimic different instrument spatial resolutions by averaging the original 3D simulation dataset at different levels (Fig.~S2a). In the zoomed-in subdomains of Fig.~S2b, we can only observe plasmoids in the first two panels while they are no longer observable in the last panel with 25$\times$25 cell average. Consistent with Fig.~2 in the  main narrative, the original dataset in Fig.~S2c can capture the new range of energy transfer in the magnetic energy spectrum with a spectral index of $-2.2$ as has been observed by NIRIS (Fig.~S2d) \cite{Abramenko2020}. While the spectrum associated with the 5$\times$5 cell averaged dataset can partially exhibit this new range of energy transfer, the spectra due to 25$\times$25 and 125$\times$125 cell averaging are no longer able to reproduce observations, consistent with the discrepancy between NIRIS and HMI observations at different spatial resolution. Here we want to reiterate that it requires further observations to rule out the coincidence of the spectral break from NIRIS observations and then to identify the underlying mechanism that leads to the spectral break.

\begin{figure*}
\begin{center}
\includegraphics[width=\columnwidth]{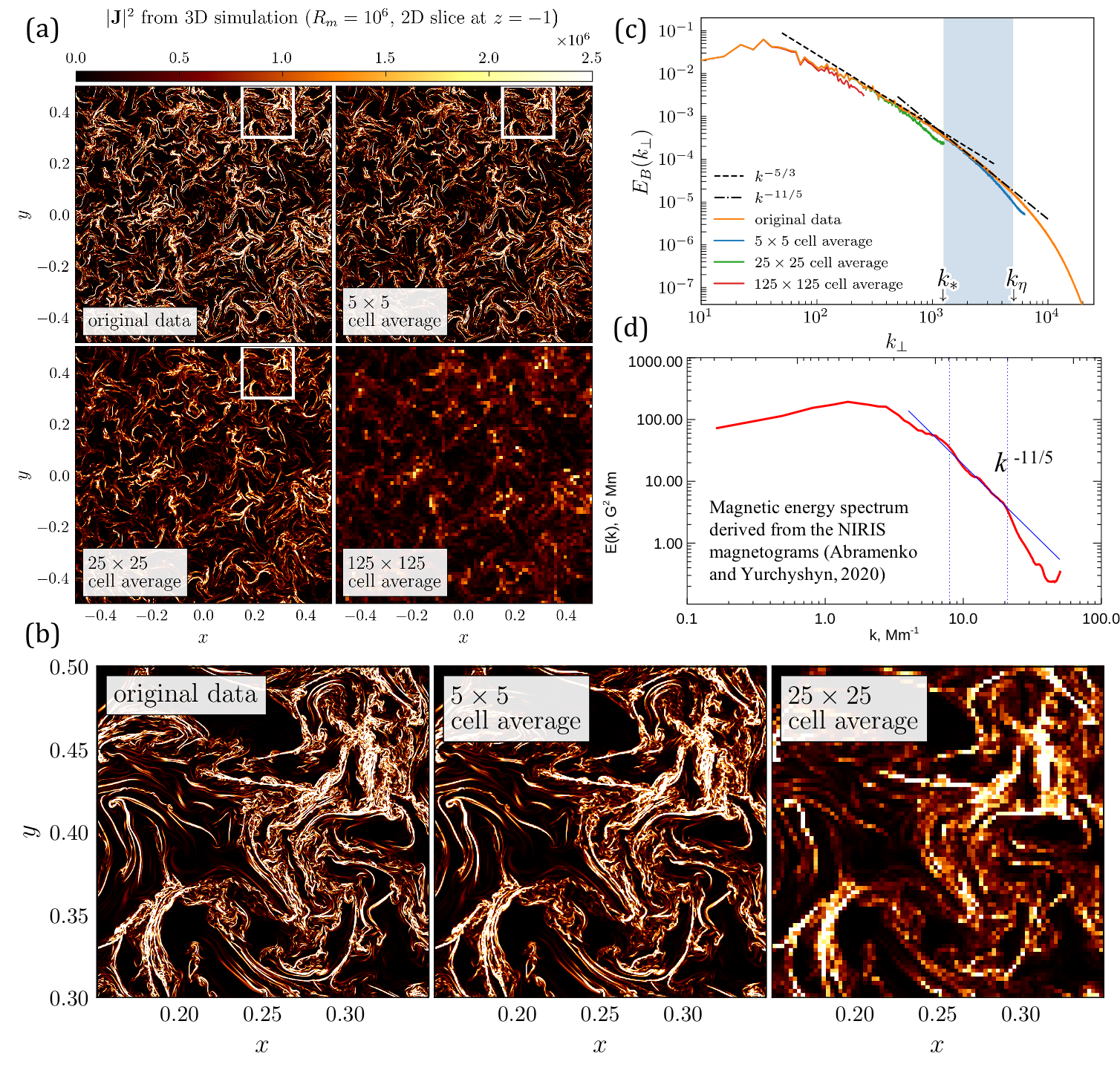}
\end{center}
\caption{{\bf Cell-averaged simulation results and comparison to spectral observations.} {\bf (a)} 2D slices of $|\mathbf{J}|^2$ (proportional to Joule heating, $\eta |\mathbf{J}|^2$, and therefore thermal emission) from the original and cell-averaged 3D datasets with $R_m = 10^6$ at $z=-1$. {\bf (b)} Zoomed-in subdomains that are highlighted in the white boxes of Panel {\bf (a)}. {\bf (c)} Magnetic energy spectra from the original and cell-averaged 3D datasets with $R_m = 10^6$. {\bf (d)} The magnetic energy spectrum derived from the high-resolution NIRIS magnetograms of the Big Bear Solar Observatory (BBSO) exhibits a new range of energy transfer with a spectral index of $k^{-11/5}$. In contrast, the spectra calculated from the 25$\times$25 and 125$\times$125 cell-averaged 3D datasets do not show this new spectral range.
}
\end{figure*}

Movie S1. \textbf{Animation of current density $|J|$.} The electric current density $|J|$ in different x-y slices through the entire 3D simulation domain for the high-$R_m$ case. \\

\newpage
\section*{Acknowledgments}

We acknowledge fruitful discussions with Muni Zhou, Silvio Cerri, Nan Liu, Manasvi Lingam, Yuxi Chen, Greg Hammett, Steven Cowley, and Adam Burrows. Resources supporting this work were provided by the NASA High-End Computing (HEC) Program through the NASA Advanced Supercomputing (NAS) Division at Ames Research Center. We also would like to acknowledge high-performance computing support from Cheyenne (doi:10.5065/D6RX99HX) provided by NCAR's CISL, sponsored by NSF, and from National Energy Research Scientific Computing Center, a DOE Office of Science user facility. \textbf{Funding:} This work was partially supported by NASA grants 80NSSC19K0621 and 80NSSC21K1326, DOE grants DE-SC0021205 and DE-SC0021254, and DOE under contract number DE-AC02-09CH11466. \textbf{Author contributions:} All of the authors made significant contributions to this work. C.D. and Y.M.H. designed the simulation for this study; C.D. carried out the simulations of MHD turbulence; C.D., L.W., Y.M.H.,  L.C., and T.A.S. analyzed the simulation results; and C.D., L.W., Y.M.H., L.C., and A.B. wrote the paper; and all authors read the paper. \textbf{Competing interests:} The authors declare no competing interests. \textbf{Data and materials availability:} All data needed to evaluate the conclusions in the paper are present in the paper and/or the Supplementary Materials.

\newpage

\end{document}